\documentclass[aps,pre,reprint,floatfix,showpacs,superscriptaddress,nofootinbib]{revtex4-1}
\usepackage{latexsym}
\usepackage{graphicx}
\usepackage{amssymb}
\usepackage{amsmath}
\usepackage{amsfonts}
\newcommand{\pa}[1]{\left(#1\right)}
\begin{document}

\title{Random Reverse Cyclic matrices and screened harmonic oscillator}

\author{Shashi C. L. Srivastava}
\email[]{shashi@vecc.gov.in}
\affiliation{RIB Group, Variable Energy Cyclotron Centre, 1/AF Bidhan nagar, Kolkata 700 064, India}

\author{Sudhir R. Jain}
\email[]{srjain@barc.gov.in}
\affiliation{Nuclear Physics Division, Bhabha Atomic Research Centre, Mumbai 400 085, India}

\begin{abstract}
We have calculated the joint probability distribution function for random reverse cyclic matrices and shown that it is related to an $N$-body exactly solvable model. We refer to this well-known model potential as a screened harmonic oscillator. The connection enables us to obtain all the correlations among the particle positions moving in a screened harmonic potential. The density of nontrivial eigenvalues of this ensemble is found to be of the Wigner form and admits a hole at the origin, in contrast to the semicircle law of the Gaussian orthogonal ensemble of random matrices. The spacing distributions assume different forms ranging from Gaussian-like to Wigner. 
\end{abstract}

\pacs{05.40.-a, 05.45.Mt, 03.65.Ge}

\date{\today}

\maketitle
\section{Introduction}
Connections between random matrix theory and exactly solvable models are very important and interesting \cite{sutherland1,sutherland,jain,jk,ajk}. It is well-known that the invariant random matrix ensembles are related to some exactly solvable many-body problems in one dimension, as was found by Calogero and Sutherland \cite{calogero,bill,moser}. In particular, the joint probability distribution function (JPDF) of the eigenvalues of random matrices shares the functional form with the probability density corresponding to the quantum ground state of the $N$-body problem. This observation is important as it allows one to obtain the correlation functions of one problem by knowing those for the other, comparing terms using a dictionary. In the same vein, even for the explanation of intermediate statistics \cite{gj}, a random matrix model was found \cite{bgs} which, in turn, was related to an $N$-particle system with an inverse-square, repulsive two-body interaction, and, an inverse-square, attractive three-body interaction \cite{jk,ajk}. Even for pseudo-Hermitian Hamiltonians (where there exists a metric $\eta $ such that $H^{\dagger} = \eta H \eta ^{-1}$), a random matrix theory can be built \cite{zj1,zj2}. The connection of this with exactly solvable models is explored in \cite{jain}. In turn, the models found in \cite{calogero,sutherland,moser} and \cite{jk,ajk} can be mapped to integrable \cite{rey} and chaotic systems \cite{jgk} for which, quite remarkably, analytically exact eigenfunctions are obtained.       

In this work, we study random, reverse cyclic matrices, that are real symmetric,
\begin{equation}
H = \begin{pmatrix} a_1 & a_2 & ...&a_N\\
a_2 & a_3 & ... & a_1\\
\vdots &  & ... & \vdots \\
a_N & a_1 & ...& \end{pmatrix}
\end{equation} 
with matrix elements chosen from an appropriate distribution function. 
Bose \textit{et al.} \cite{Bose02} derived the limiting spectral distribution for reverse-cyclic matrices, but the JPDF and the spacing distribution function remain open problems. In fact it will be interesting to see how the special symmetric matrices having a very small degree of freedom (only $N$ in this case), differ from the results known for their counterparts having a full degree of freedom [\textit{i.e.} $\frac{N(N+1)}{2}$]. We have recently obtained the JPDF for the cyclic matrices, which forms another example in which the degree of freedom of the matrix is constrained (again only $N$) in an asymmetric matrix for which a spectrum of a spacing distribution from Gaussian, to Wigner and not-so-Wigner type \cite{Jain08} is obtained. The results were also used to study a random walk problem on a one-dimensional disordered lattice \cite{Mani11} where the evolution matrix is cyclic. On the one hand, there is a vast literature about different results for random cyclic  matrices in literature (see \cite{Bose09}\cite{Meckes09} \textit{etc.}). However, the same is not true for random reverse-cyclic matrices. Interestingly, reverse cyclic matrices appear (albeit with the name reverse circulant and retro-circulant) in models for particle masses, flavour mixing, and CP violation. Here families of particles can be shown to emerge by a spontaneous breakdown of discrete ${\bf Z}_6$ chiral symmetry, by the Higgs sector \cite{adler}. The presence of reverse-cyclic matrices is due to $S_3$ cyclic permutation symmetry of the Lagrangian. Quoting Adler, ``...in the limit of $S_3$ cyclic permutation symmetry, we shall find that the fermion mass matrices in both the three and six doublet models are retrocirculants..."\cite{adler}. In another instance, while exploring whether discrete flavor symmetry $S_3$ can explain the pattern of neutrino masses and mixings, reverse-cyclic matrices (again referred to as retro circulant) have been used as a perturbation matrix\cite{Dev11}. It was also shown in \cite{Dev11}  that after third order perturbation, neutrino mixing depends only on perturbation parameter, consistent with experimental data. One may speculate that the background and statistical errors may make these matrices random.

In the following, we present the JPDF for the random reverse-cyclic matrices, and, an exactly solvable model related to this problem. The form of potential (for a single particle) has been discussed in quite a few physical situations. It has been interpreted as a screened, two-dimensional isotropic harmonic oscillator in a different context \cite{Davidson32}. It has found use in explaining roto-vibrational states in the case of diatomic molecules by considering a five-dimensional version of the Davidson oscillator \cite{Rowe05}, and in a different context of dynamical symmetries \cite{wu2000} and uncertainty relations \cite{Patil07}. In the context of many-body physics, we might imagine the above Hamiltonian describing bosons in a harmonic trap ($r^2$ term), interacting via a  dipolar electric field ($r^{-2}$ term). 

We collect some known results related to the eigen-decomposition of reverse-cyclic matrices. A known eigen-decomposition becomes a very advantageous tool, to derive the joint probability distribution function for eigenvalues. Karner \textit{et al.} \cite{Karner03} have shown that the eigen-decomposition for an odd-dimensional reverse cyclic  matrix  is given by
\begin{eqnarray}\label{eq:decomp}
H &=& F^\dagger \begin{pmatrix}1 & 0\\ 0 & R\end{pmatrix} \Lambda \begin{pmatrix}1 & 0\\ 0 & R^\dagger \end{pmatrix} F\\ \nonumber
\Lambda &=& (E_1, |E_2|,\ldots, |E_{(n-1)/2}|,-|E_{(n-1)/2}|,\ldots, -|E_2|)
\end{eqnarray} 

\begin{equation}\label{eq:four}
F_{r,s}(n)=\frac{1}{\sqrt n}  e^{2\pi  i(r-1)(s-1)/n};~~~~r,s = 1,2,\ldots,n.
\end{equation}

\begin{eqnarray}\label{eq:ortho}
R &:=&\begin{pmatrix}E^\dagger & i E^\dagger \hat{I_k}\\ \hat{I_k} E & -i\hat{I_k} E\hat{I_k} \end{pmatrix}\in \mathbb{C}^{2k\times 2k} \\ \nonumber
&~& \mbox{with}~~~E=\frac{1}{\sqrt 2} \mbox{diag}\left(e^{i\phi _1/2},\ldots,e^{i\phi _k/2}\right)
\end{eqnarray}
$\hat{I_k}$ is an anti-diagonal identity matrix, and $0\leq \phi_j<2\pi$. The eigen-decomposition for an even-dimensional reverse cyclic matrix takes following form,
\begin{eqnarray}\label{eq:decompeven}
H &=& F^\dagger \begin{pmatrix}1 & 0\\ 0 & R\end{pmatrix} \Lambda_1 \begin{pmatrix}1 & 0\\ 0 & R^\dagger \end{pmatrix} F\\ \nonumber
\Lambda_1 &=& (E_1, |E_2|,\ldots, |E_{(n-2)/2}|,E_{n/2},-|E_{(n-2)/2}|,\ldots, -|E_2|)
\end{eqnarray} 
with
\begin{eqnarray}\label{eq:orthoeven}
R &:=&\begin{pmatrix}E^\dagger & 0& i E^\dagger \hat{I_k}\\0 & 1 &0 \\  \hat{I_k} E & 0& -i\hat{I_k} E\hat{I_k} \end{pmatrix}\in \mathbb{C}^{2k+1\times 2k+1}
\end{eqnarray}
 
\section{JPDF, spacing distribution and discussion} 
Consider an ensemble of reverse cyclic (RC) matrices, drawn from a Wishart distribution,
\begin{equation}\label{eq:ph}
P(H)\sim \exp\left(-A \mbox{Tr}(H^\dag H\right)).
\end{equation}
 
\begin{figure}
\begin{center}
\includegraphics[width=1.0\columnwidth, height=0.7\columnwidth]{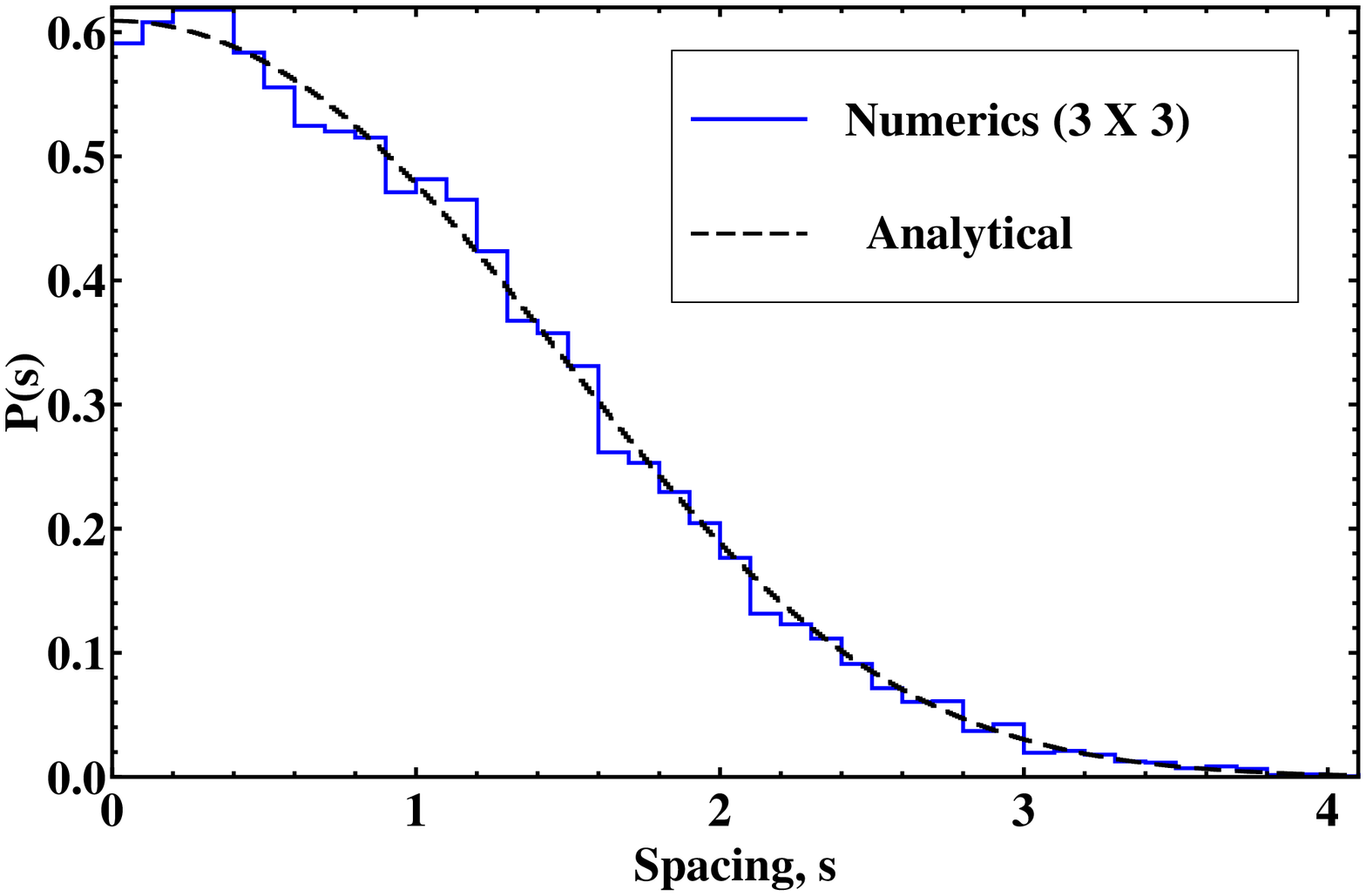} 
\includegraphics[width=1.0\columnwidth, height=0.7\columnwidth]{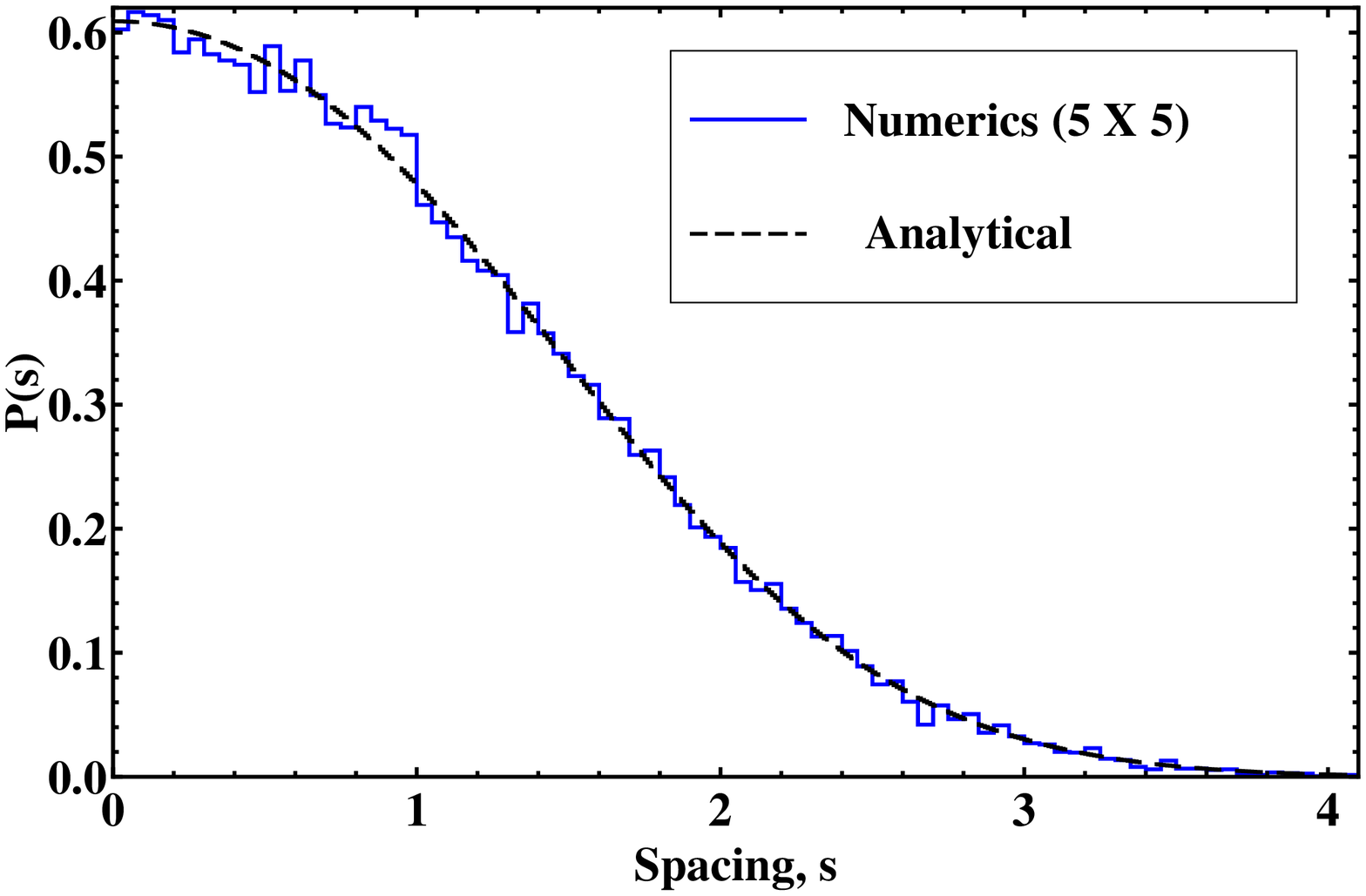} 
\includegraphics[width=1.0\columnwidth, height=0.7\columnwidth]{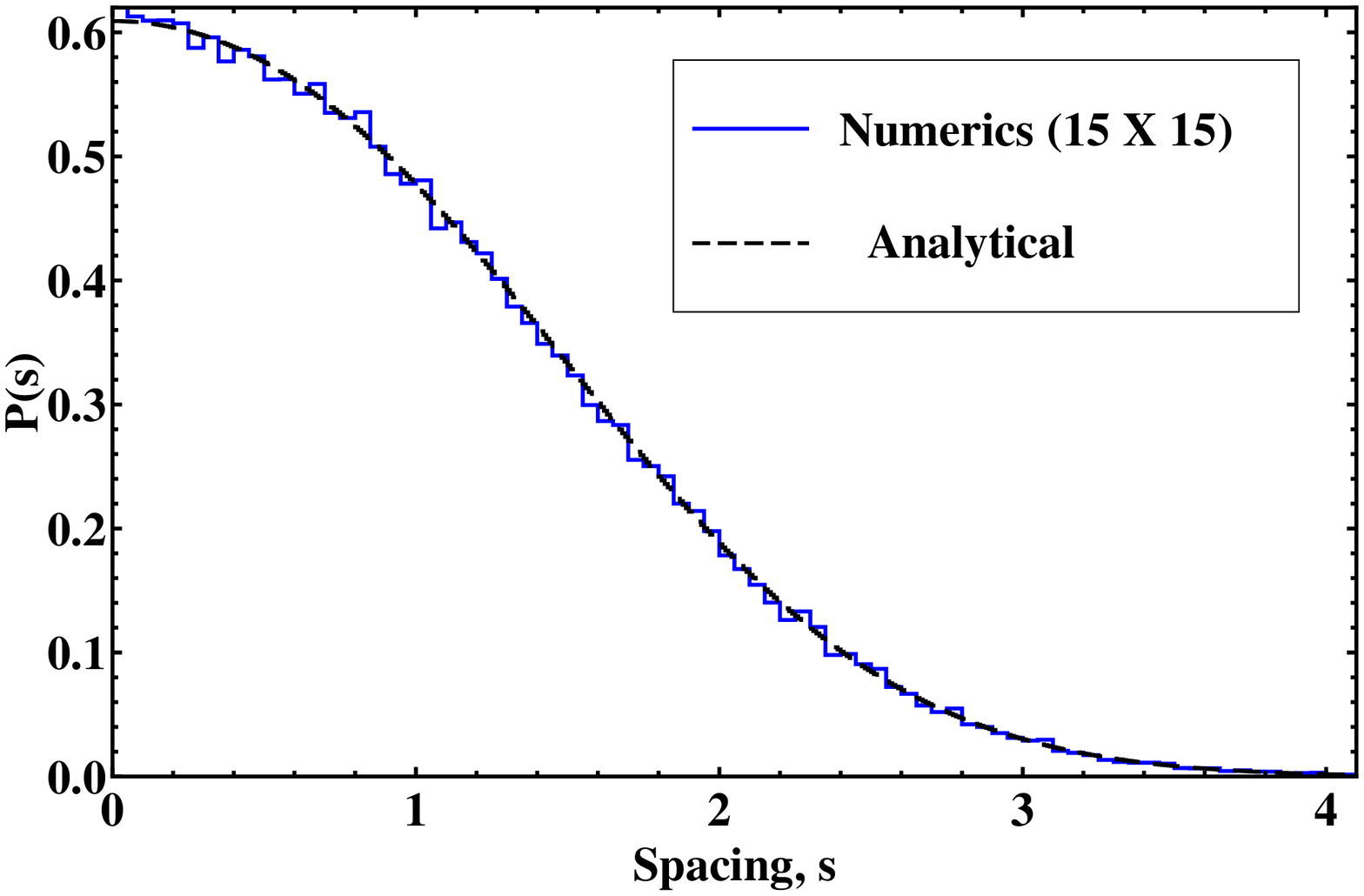} 
\end{center}
\caption{(Color online)Spacing distribution $s_{12}$ for an ensemble of 20000 reverse cyclic matrices of size $3\times 3$,  $5\times 5$ and $15\times 15$  is compared with the analytical form (eq.\ref{eq:spacing3}).}
 \label{fig:spacing3}
\end{figure}

\begin{figure}
\begin{center}
\includegraphics[width=1.0\columnwidth, height=0.7\columnwidth]{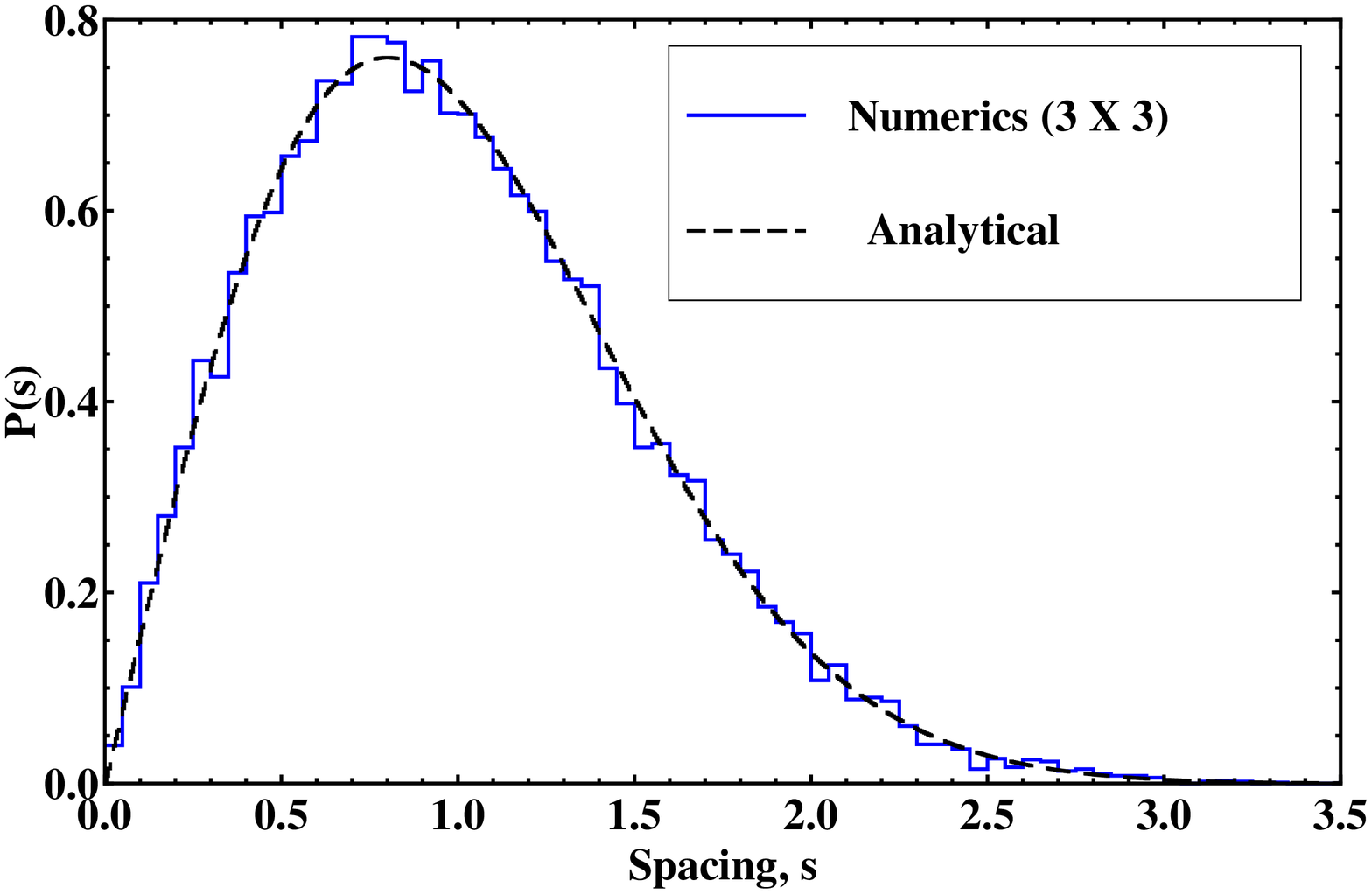}\\
\includegraphics[width=1.0\columnwidth, height=0.7\columnwidth]{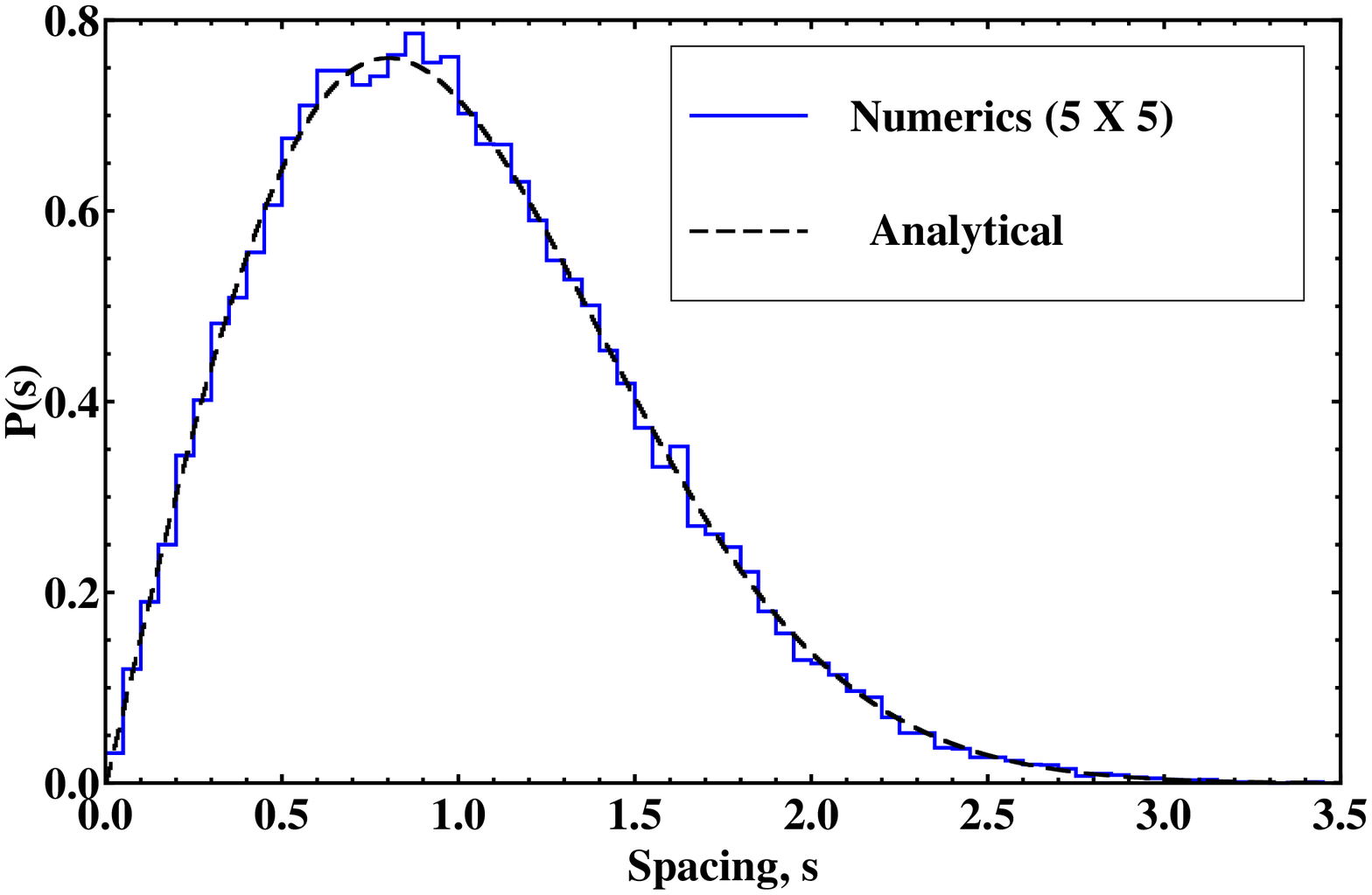}
\includegraphics[width=1.0\columnwidth, height=0.7\columnwidth]{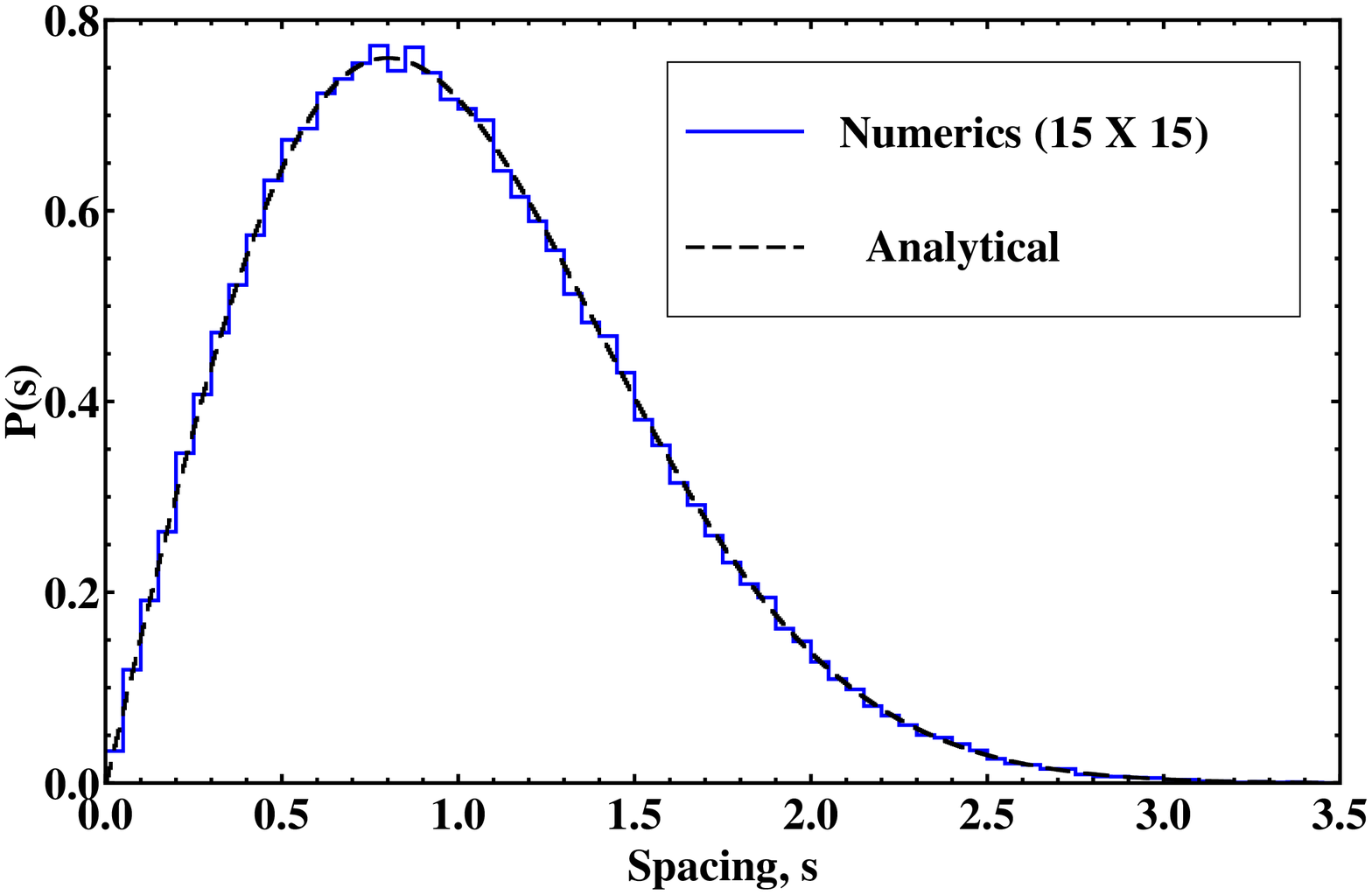}
\end{center}
\caption{(Color online)Spacing distribution $s_{23}$ for an ensemble of 20000 reverse cyclic matrices of size $3\times 3$, $5\times 5$ and $15\times 15$ is compared with the analytical form (eq.\ref{eq:spacing3rep}).}
 \label{fig:spacing3rep}
\end{figure}

Let us start with the simplest case, namely an ensemble of $3\times 3$ reverse cyclic matrices, 
\begin{equation}
H = \begin{pmatrix} a & b & c\\
b & c & a\\
c & a & b \end{pmatrix}.
\end{equation}
The JPDF in matrix space will be given by, using (\ref{eq:ph}), by
\begin{equation}\label{eq:ph3}
P(a, b, c) = \left(\frac{3 A}{\pi}\right)^{(3/2)} \exp [-3 A \left(a^2+b^2+c^2\right)].
\end{equation}
From Eq. (\ref{eq:decomp}), we can diagonalize  $H$ and it is also clear that there are only $(n+1)/2$ independent eigenvalues for odd-dimensional matrices. For the $3 \times 3$ case, the explicit form of $R$ is 
\begin{equation}
R = \begin{pmatrix}1 & 0 & 0\\ 
     0 & \frac{1}{\sqrt 2} \exp(-i \theta/2) & \frac{i}{\sqrt 2} \exp(-i \theta/2)\\
     0 & \frac{1}{\sqrt 2} \exp(i \theta/2) & -\frac{i}{\sqrt 2} \exp(i \theta/2) \end{pmatrix}.
\end{equation}
It takes a simple algebra then to show that
\begin{eqnarray}\label{eq:abc_eig}
\nonumber
a&=&\frac{1}{3} (E_1+2 |E_2|\cos \theta) \\
b&=&\frac{1}{3} \left( E_1- |E_2| \left(\cos \theta +\sqrt 3 \sin \theta \right)\right) \\ \nonumber
c&=&\frac{1}{3} \left(E_1- |E_2|\left(\cos \theta -\sqrt 3 \sin \theta \right)\right).
\end{eqnarray}
Using (\ref{eq:abc_eig}) in (\ref{eq:ph3}), we can find the jpdf for eigenvalues and an independent parameter $\theta$ coming from the eigenvector. Note that in $H$, the independent parameters are three in number, namely $a,b~\mbox{and}~c$; while in the eigen-decomposition, we have $E_1, E_2, \theta$. The Jacobian for the transformation (\ref{eq:abc_eig}) is given by $\frac{2 \left|E_2\right|}{3 \sqrt 3}$.
The jpdf for eigenvalues is 
\begin{eqnarray}
\nonumber
P(E_1,|E_2|,\theta) &=& \frac{2 |E_2|}{3 \sqrt 3} \left(\frac{3 A}{\pi}\right)^{(3/2)} \exp [-A \left(E_1^2 + 2 E_2^2\right)]\\ \nonumber
 &~& ~~\mbox{where}~ E_1 \in (-\infty,\infty), |E_2|\in [0,\infty), \\  
 &~& ~~~ \theta \in [0,2\pi).
\end{eqnarray}
Notice that the domain of $|E_2|$ is $[0,\infty )$, and that the function on the right hand side is an even function of $E_2$, thus we can rewrite the JPDF after an integration over $\theta$ in the following form,
\begin{equation}
P(E_1,E_2) = 2\pi\frac{|E_2|}{3 \sqrt 3} \left(\frac{3 A}{\pi}\right)^{(3/2)} \exp(-A \left(E_1^2 + 2 E_2^2\right)).
\end{equation}
The density of $E_1$ (the trivial eigenvalue) \cite{footnote} comes out to be Gaussian as expected because of $E_1$ being a sum of Gaussians. On the other hand, the density of non-trivial eigenvalue $E_2$ is given by (\ref{eq:den3}).
\begin{equation}\label{eq:den3}
\rho(E) = 2 A |E| \exp(-2 A E^2)
\end{equation}
\begin{figure}
\begin{center}
\includegraphics[width=1.0\columnwidth, height=0.7\columnwidth]{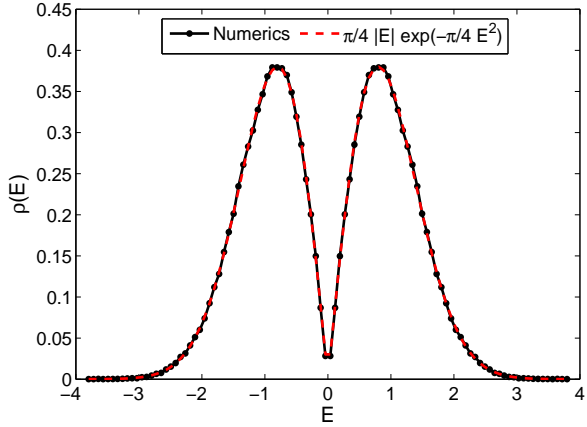}
\end{center}
\caption{(Color online)Normalized density of non-trivial eigenvalues for an ensemble of 20000 reverse cyclic matrices of size $15\times 15$  is compared with the analytical form. The density is normalized such that averaged density for positive eigenvalues is 1/2.}
 \label{fig:density}
\end{figure}
Also, due to product structure of the JPDF, the density of non-trivial eigenvalue will remain the same for higher-dimensional matrices. The presence of $|E|$ ensures that there are no non-trivial eigenvalues present at origin while they increase linearly along both the positive and negative real axis. It is as if there is a hole in the density of non-trivial eigenvalues (see Fig.\ref{fig:density}). This has been independently derived by Bose \textit{et al.} \cite{Bose02} without obtaining the JPDF.  Also notice that, it is the limiting distribution in the case of \cite{Bose02} while here it is an exact result for any dimension (matrix). The spacing distribution  between $E_1, E_2$ can now be calculated as
\begin{eqnarray}\label{eq:spacing3}
\nonumber
P(s_{12})&=& \int_{-\infty}^{\infty} \int_{-\infty}^{\infty} P(E_1, E_2)\delta(s_{12}-|E_1 - E_2|)\\ \nonumber
         &=&  \frac{ 12 \sqrt A e^{-A s_{12}^2}}{9 \sqrt \pi} \\ 
         &~& + \frac{4 A e^{-2 A \frac{s_{12}^2}{3}} \sqrt{3 \pi} s_{12} \mbox{Erf}\left(\sqrt{\frac{ A}{3}} s_{12}\right) }{9 \sqrt \pi}.
\end{eqnarray}
The value of $A$ can be chosen so that $\int_0^{\infty} s_{12} P(s_{12}) ds_{12} =1$. A numerical histogram is compared with (\ref{eq:spacing3}) in Fig.\ref{fig:spacing3}.
One could think of spacing between the second and third eigenvalue of $H$, but due to their special form as $|E_2|$ and $-|E_2|$, it is simply given by $s_{23} = 2|E_2|$, so the spacing distribution as expected is very similar to the density of $|E_2|$ and is given by (\ref{eq:spacing3rep}):
\begin{equation}\label{eq:spacing3rep}
P(s_{23}) = A s_{23} e^{-\frac{A}{2} s_{23}^2}.
\end{equation}
Again, the value of $A$  is chosen such that $\int_0^{\infty} s_{23} P(s_{23}) ds_{23} =1$, which turns out to be $\pi/2$. A comparison with the numerical data is shown in Fig. \ref{fig:spacing3rep}.

\begin{figure}
\begin{center}
\includegraphics[width=1.0\columnwidth, height=0.7\columnwidth]{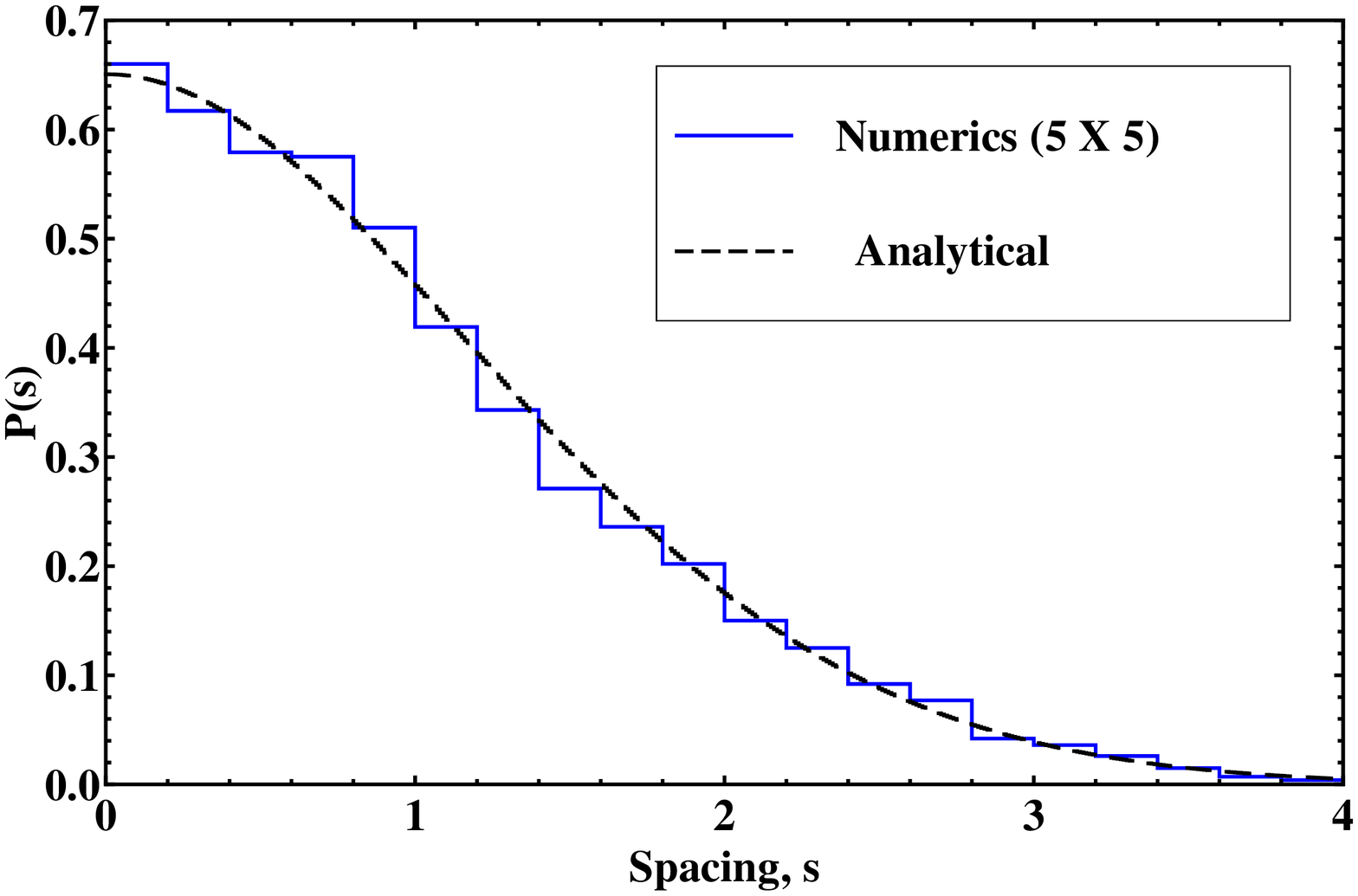}\\
\includegraphics[width=1.0\columnwidth, height=0.7\columnwidth]{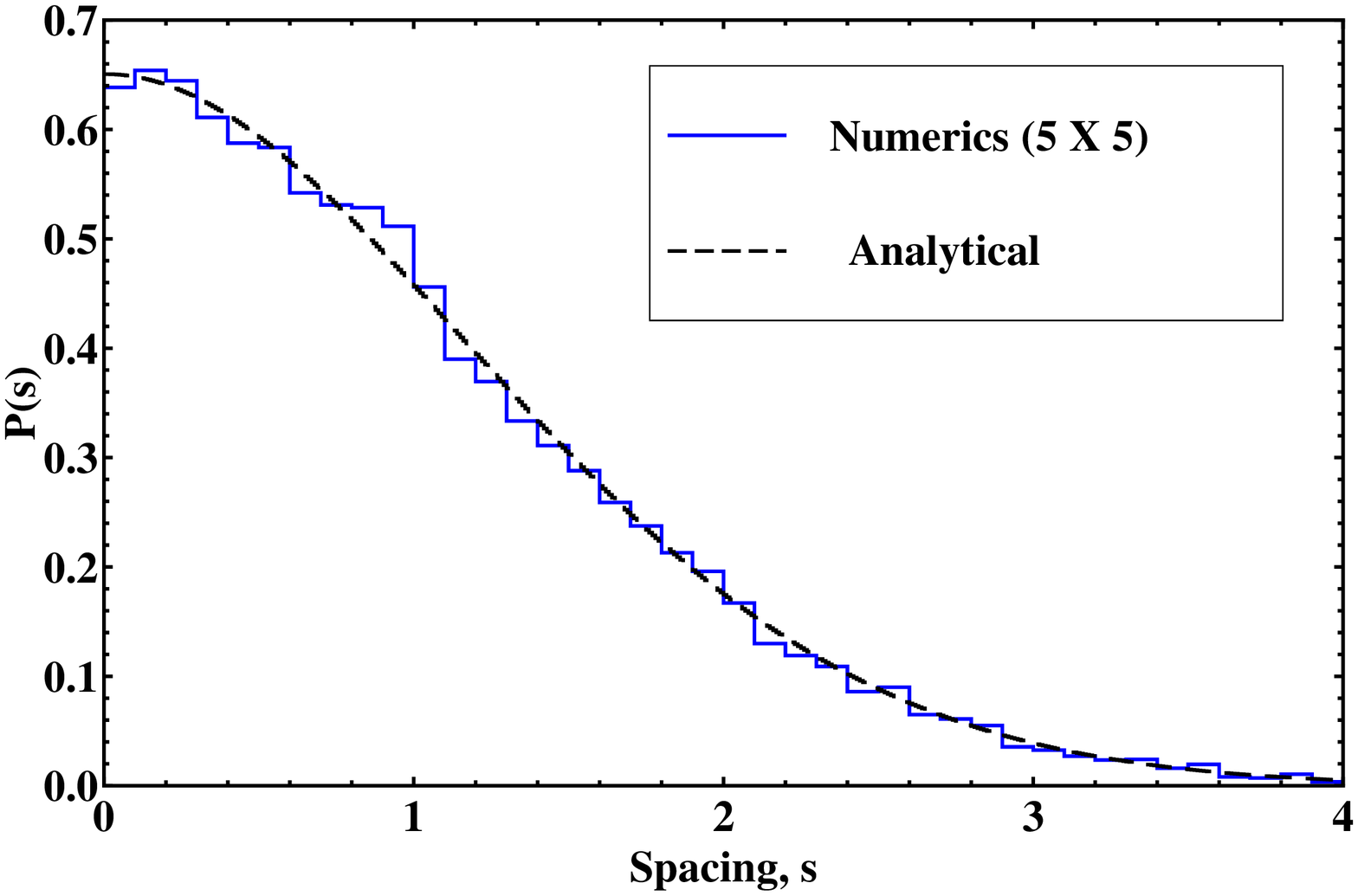}
\includegraphics[width=1.0\columnwidth, height=0.7\columnwidth]{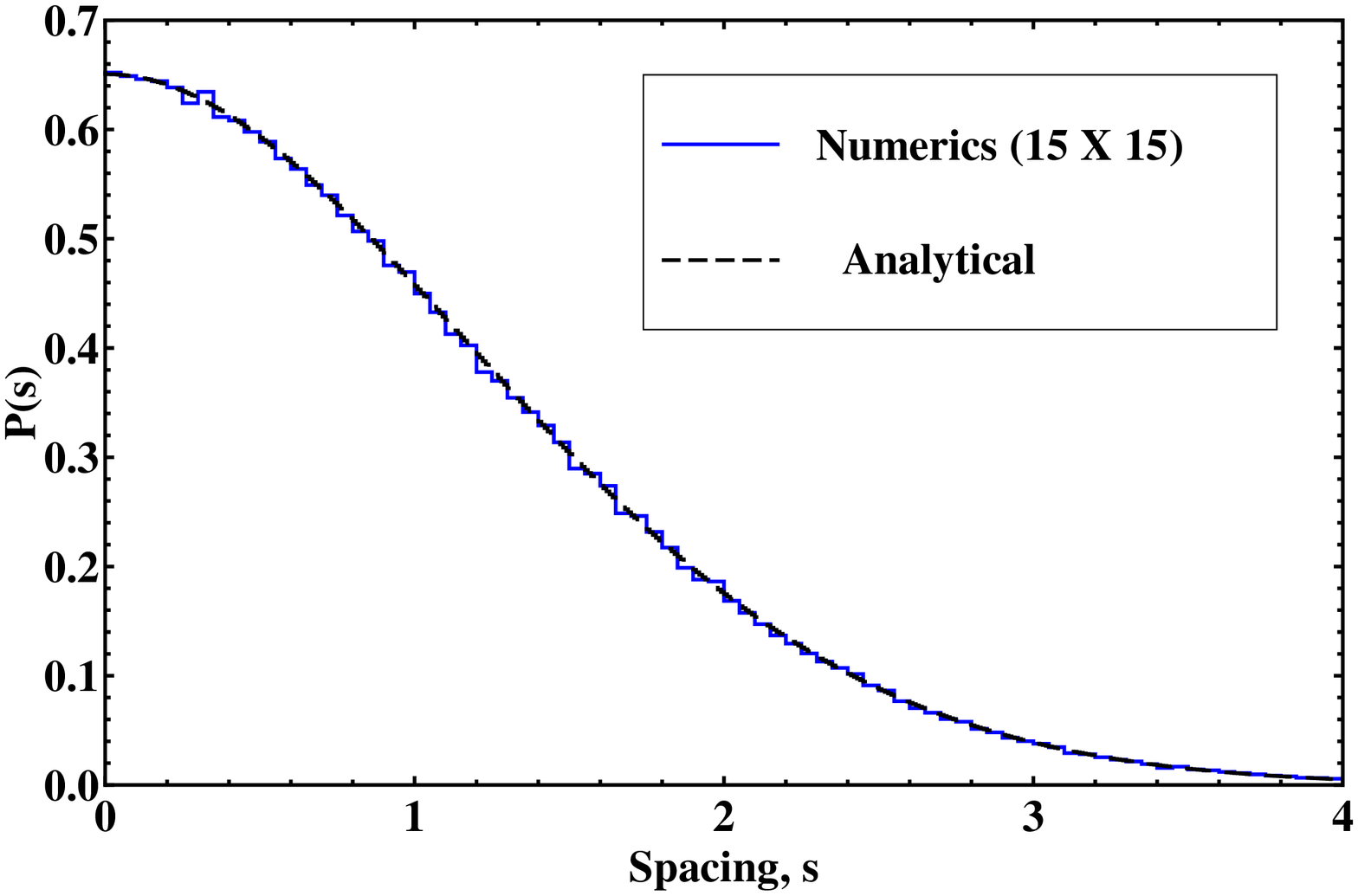}
\end{center}
\caption{(Color online)Spacing distribution $s_{pp}$ for an ensemble of 5000 and 20000 reverse cyclic matrices of size $5\times 5$ and $15\times 15$ is compared with the analytical form (eq.\ref{eq:spac5positive}).}  
 \label{fig:spac5positive}
\end{figure}

In the case of $5\times 5$, a similar procedure will give the JPDF as in (\ref{eq:jpdf5}) with $E_i \in (-\infty,\infty)$ and $\theta_i \in [0,2\pi)$:
\begin{eqnarray}\label{eq:jpdf5}
\nonumber
P(E_1,E_2,E_3,\theta_1,\theta_2 ) &=& \frac{|E_2| |E_3|}{25 \sqrt{5}} \left( \frac{5 A}{\pi}\right)^{5/2} \times \\ &~&\exp [-A (E_1^2 + 2E_2^2 + 2E_3^2)].
\end{eqnarray}
The density of $E_i$s and the spacing distribution for the cases appearing in $3\times 3$ reverse cyclic matrices remain the same. There is an additional spacing possible, namely between two positive $|E_2|$ and $|E_3|$. Let us denote this by $s_{pp}$. Its distribution is 
\begin{eqnarray}\label{eq:spac5positive}
\nonumber
P(s_{pp})&=& \int_{-\infty}^\infty \int_0^{\infty} \int_0^{\infty} 4 P(E_1, E_2, E_3)\delta(s_{pp}-|E_2 - E_3|)\\ \nonumber
         &=&  A s_{pp} e^{-2 A s_{pp}^2}   \\ 
         &~& -\frac{1}{2}\sqrt{\pi A} e^{-A s_{pp}^2} (-1 + 2 A s_{pp}^2) \mbox{Erfc}(\sqrt{A} s_{pp}).
\end{eqnarray}
The area under this distribution is 1/2. Taking care of the domains of $|E_2|$ and $|E_3|$, and accounting for the spacing between these and between $-|E_2|$ and $-|E_3|$, we obtain the correctly normalized distribution.  This can be seen to be in agreement with the numerical data (see Fig. \ref{fig:spac5positive}).  This same distribution [Eq. \ref{eq:spac5positive}]has been compared with the distribution of spacings among all positive eigenvalues except the Gaussian distributed one of an ensemble of higher-dimensional reverse-cyclic matrices (\textit{e.g.} $15 \times 15$). The agreement is good.
For the ($2n+1$) dimensional reverse-cyclic matrix JPDF is a straightforward generalization of (\ref{eq:jpdf5}) and is given by (\ref{eq:jpdfn}).
\begin{eqnarray}\label{eq:jpdfn}
\nonumber
P(E_1,E_2,\ldots,E_{2n+1},\theta_1,\ldots,\theta_{2n}) = \left( \frac{A}{\pi}\right)^{(2n+1)/2} \\  
|E_2|\ldots|E_{n+1}| \exp [-A (E_1^2 + 2\sum_{i=2}^{n+1}E_i^2)].
\end{eqnarray}
This JPDF can be understood as follows. Clearly, the nature of the first (trivial) eigenvalue is very different from the others (nontrivial), and its distribution will be Gaussian. We focus on the rest of the eigenvalues. The diagonalizing equation, $H=O\Lambda O^T$ where $O = F^\dagger R$ is an orthogonal matrix \cite{Karner03}, has the correct number of independent parameters. For a $(2n+1)$-dimensional matrix $H$, the left hand side has only $(2n+1)$ independent variables while the right-hand side has $(n+1)$ independent eigenvalues with $n$ angle variables in $O$. As $dH$ will contain $(2n+1)$ independent differentials, so a multiplication of $H$ with a scalar $a$ will satisfy $d(aH) = a^{2n+1} dH$. Now, $(n+1)$ of them will be absorbed in the scaling of measure $d\Lambda$[as independent eigenvalues are $(n+1)$]. Hence, from the scaling property of $d(aH)$, $dH$ will be a homogeneous polynomial of degree $n$\cite{Forrestor}. Our prototype examples for $n=3$ and 5 has shown that they vanish linearly as eigenvalues approach the origin,  hence the polynomial in the eigenvalues is necessarily  proportional to $|E_2|\ldots|E_{n+1}|$. The even case is not very different from the odd one, except that $E_{n/2 +1}$ appears along with $E_1$, the rest being the same as that in (\ref{eq:jpdfn}).
\section{Screened harmonic oscillator and JPDF}
Now we show the exactly solvable $n$-body problem, the ground-state wave function of which is such that the probability density has the same mathematical form as (\ref{eq:jpdfn}).  
It can be verified that (\ref{eq:jpdfn}) corresponds to $|\Psi (x_1, x_2, ..., x_n)|^2$ where $\Psi (x_1, x_2, ..., x_n)$ is the ground state wave function with eigenvalue $(4n-3) A$ of the $n$-body problem with the Hamiltonian:
\begin{equation}\label{eq:nbodyh}
\mathcal{H}(x_1, x_2, ..., x_n )= - \nabla^2 + \left[ A^2 x_1^2 +\sum_{i=2}^n \pa{4 A^2 x_i^2 - \frac{1}{4x_i^2}}\right].
\end{equation}
To illustrate that this is so, let us verify for $n=2$. This will also be sufficient for general $n$ due to the  identical form of separable $\mathcal{H}$. As we need to take double derivatives of the wave function, it will be prudent to replace $|E_i|$s by $\sqrt{E_i^2}$. Hence, $\Psi (x_1, x_2) = c \sqrt{\sqrt{x_2^2}}\exp\pa{-A/2\pa{x_1^2 + 2x_2^2}}$, the JPDF for a corresponding three-dimensional reverse-cyclic matrix,
\begin{eqnarray*}
\frac{\partial^2}{\partial x_1^2} \Psi (x_1, x_2) = A \pa{-1 + A x_1^2} \Psi (x_1, x_2) \\
\frac{\partial^2}{\partial x_2^2} \Psi (x_1, x_2) = \frac{\pa{-1 + 16 A x_2^2 \pa{-1 + A x_2^2}}}{4} \Psi (x_1, x_2)\\
\pa{-\frac{\partial^2}{\partial x_1^2}-\frac{\partial^2}{\partial x_2^2} + A^2 x_1^2 + 4 A^2 x_2^2 - 1/(4 x_2^2)} \Psi (x_1, x_2) \\ ~~~~~= 5 A \Psi (x_1, x_2)
\end{eqnarray*}
This proves our assertion.

This system has been the subject of a lot of work, initiated by Perelomov \cite{perelomov}. The only potential that can be added to a harmonic interaction is $2a/x^2$ if we want to successfully construct the creation and annihilation operators for the above model \cite{hoppe}. This work relates this well-known model to a random matrix theory for reverse-cyclic matrices, which constitutes a remarkable addition to the known connections along similar lines.

The linear level repulsion obtained here has its origin in the product of the absolute value of the eigenvalues in the JPDF. This is reflected in the interaction among eigenvalues if we write the JPDF as a partition function for an $n$-particle system. This interaction, in the context of random matrices is the Coulomb interaction in two dimensions. In contrast, the case of random-cyclic matrices \cite{Jain08} has a JPDF which is just the exponential containing a sum of the square of the modulus of the complex eigenvalues. The eigenvalues are in a plane, and the level repulsion comes out as a Rayleigh distribution for the Poisson process on a plane,  which has the same functional form as Wigner's spacing distribution for the orthogonal ensemble. Thus, we have a very interesting situation for the random reverse-cyclic and random cyclic matrices in that we obtain the same formula for the spacing distribution but the origin is different.   

\section{Summary}
In summary, we have shown that reverse-cyclic matrices though a subset of symmetric matrices have an unusual density and spacing distribution. In contrast to semi-circle density, this ensemble admits a density with a hole at the origin. Again, the spacing distribution has a variety ranging from Gaussian-looking distributions to Wigner type distributions.  We also observed that the JPDF is just the square of the modulus of the ground-state eigenfunction of an exactly solvable many-body Hamiltonian in one dimension, of a screened harmonic oscillator potential. Hence the correlations between the different particles in the potential will be the same as that derived from the joint probability distribution function for the random matrix theory.

\begin{acknowledgments}
The authors would like to thank Arul Lakshminarayan, Indian Institute of Technology, Madras for useful discussions.
\end{acknowledgments}


\begin{thebibliography}{99}
\bibitem{footnote} One eigenvalue of reverse cyclic matrices are just sum of elements due to permutation symmetry, this will be called as trivial eigenvalue.
\end{thebibliography}
\end{document}